# Drivers of social influence in the Twitter migration to Mastodon

**Lucio La Cava**[1,*], **Luca Maria Aiello**[2,3], **and Andrea Tagarelli**[1]

[1]University of Calabria, Rende, Italy
[2]IT University of Copenhagen, Denmark
[3]Pioneer Centre for AI, Denmark
[*]lucio.lacava@dimes.unical.it

## ABSTRACT

The migration of Twitter users to Mastodon following Elon Musk's acquisition presents a unique opportunity to study collective behavior and gain insights into the drivers of coordinated behavior in online media. We analyzed the social network and the public conversations of about 75,000 migrated users and observed that the temporal trace of their migrations is compatible with a phenomenon of social influence, as described by a compartmental epidemic model of information diffusion. Drawing from prior research on behavioral change, we delved into the factors that account for variations of the effectiveness of the influence process across different Twitter communities. Communities in which the influence process unfolded more rapidly exhibit lower density of social connections, higher levels of signaled commitment to migrating, and more emphasis on shared identity and exchange of factual knowledge in the community discussion. These factors account collectively for 57% of the variance in the observed data. Our results highlight the joint importance of network structure, commitment, and psycho-linguistic aspects of social interactions in characterizing grassroots collective action, and contribute to deepen our understanding of the mechanisms that drive processes of behavior change of online groups.

## Introduction

After years of steady growth, popular social media are experiencing shifts in user engagement. In Twitter, such a shift has been especially abrupt after business magnate Elon Musk purchased the platform on October $26^{th}$ 2022. The acquisition itself, as well as several controversial management decisions taken by Musk shortly after[1] (including massive layoffs, the suspension of some journalists' accounts, and the discontinuation of free API access) threw Twitter at the center of a media storm and caused abrupt changes in the typical platform activity,[2] motivating many users to seek substitute services to migrate to.

Decentralized Online Social Networks (DOSNs) have experienced significant growth in recent years, capturing the attention of mainstream social media users.[3] Among these networks, *Mastodon* has emerged as the leading decentralized alternative to Twitter.[4–7] Similar to email services, Mastodon allows communities to independently manage their own "instances" (i.e., servers) and connect with others through a federated approach facilitated by a common protocol. Following Elon Musk's takeover, Twitter users advocating for a transition to Mastodon promoted the #TwitterMigration movement, resulting in a rapid surge of registration requests on Mastodon instances.

This mass exodus from Twitter represents one of the largest digital migrations in the history of the Social Web and a unique example of collective behavioral change that is documented through large-scale digital traces, that can thus be studied quantitatively and at scale. Moreover, this phenomenon exhibits two uncommon properties that render it especially interesting from the perspective of behavioral change studies.[8] First, despite being prompted by external circumstances, the migration unfolded organically within Twitter, with users engaging in discussions and potentially influencing their peers by signaling their intention to migrate. Second, transitioning to a different social platform entails practical and psychological costs associated with changing habits,[9] as well as a social cost associated with adopting a behavior that deviates from mainstream norms;[10] these characteristics are shared with other grassroots processes of behavioral change that are generally desirable for human societies,[11–13] which further speaks to the significance of studying this migration.

Early studies have taken initial steps in characterizing the #TwitterMigration phenomenon primarily in terms of user activity, revealing that the majority of migrated users congregated on a few Mastodon instances,[14] while also continuing to post content on Twitter.[15] However, the underlying *drivers* motivating Twitter users to migrate to Mastodon remain largely unexplored, and this work is a first attempt to fill this gap.

We hypothesize that the migration was partly determined by social influence. Drawing inspiration from previous studies,[16,17] we employed a compartmental model of information diffusion to describe the phenomenon at a macroscopic level, and examined whether the temporal pattern of migration is compatible with the typical dynamics of information contagion. We then shifted our analysis to the mesoscopic level and investigated different communities on Twitter to identify the factors that account for variations in the effectiveness of the influence process. Crucially, the factors we considered are rooted in three branches of interdisciplinary research on behavior change, which led us to formulate three distinct hypotheses regarding the *structure* of the social network, the *commitment* of community members, and their *language use*.

First, we hypothesize that the structural characteristics of the Twitter social network correlate with the rate of behavioral contagion across communities. This hypothesis is motivated by sociological and network science studies highlighting how structural attributes of social graphs, such as group size, cohesiveness, and the presence of influential authorities, can influence the dynamics of information diffusion and behavioral change within social networks.[18] Our second hypothesis draws on controlled experiments demonstrating that a committed minority of individuals can trigger tipping points in opinion formation.[19,20] Consequently, we examine whether communities exhibiting higher commitment to the #TwitterMigration discussion also displayed faster influence processes. Our third and final hypothesis builds upon prior research in social psychology, which has established connections between psycho-linguistic aspects of social interactions and various outcomes related to consensus, agreement, and coordination.[21–25] Leveraging recent advances in natural language processing, we extracted high-level language dimensions that convey specific social intents.[26] Here, we hypothesize that communities engaging in conversations rich with positive social intent, such as knowledge exchange, expressions of trust, and identity markers, exhibit a more process of social influence.

Our findings reveal that communities where the peer influence to migrate appears to be more effective exhibit lower density of social connections, higher commitment to the discussion, and frequent emphasis on shared *identity* and exchanges of factual *knowledge*. Collectively, these factors explain more than half of the observed variance in the data.

Our study departs from previous research on online migrations. Existing studies that aimed to characterize user transitions from one digital platform to another have either predominantly resorted to qualitative approaches[27–30] or relied on quantitative analyses conducted on small-scale datasets.[31–33] A few studies have explored user migrations at scale in blockchain-based online social networks, describing the effects of the migration on the structure of the social graph,[34,35] and finding that network density is a predictor of migration.[36] Recently, a distinct line of research has emerged, focusing on user migrations prompted by *deplatforming*,[37] which refers to the removal of accounts by social media administrators due to user engagement in toxic, offensive, or abusive activities. Studies in this domain have primarily examined the efficacy of deplatforming in disbanding groups of deviant users and reducing the prevalence of toxic interactions online, yet the findings indicate limited success in achieving these objectives.[38–41] Migration resulting from deplatforming fundamentally differs in nature from the phenomenon under investigation in our study, as it is coerced rather than organic.

## Results

### Following the Migration

We collected approximately 2M tweets related to the #TwitterMigration phenomenon, spanning from October 26th, 2022, to January 19th, 2023. Our analysis focused on a subset of over 1.3M tweets written in English, contributed by approximately 500K unique authors. To determine which of these authors had migrated to Mastodon, we leveraged self-disclosed information provided by the authors themselves. During the migration process, numerous users openly shared their Mastodon handles on Twitter to advertise their presence on the new platform. To identify these individuals, we conducted a comprehensive search within the usernames, public profiles, and tweets of all users in our dataset, specifically seeking mentions of Mastodon handles. We found around 75K valid Mastodon handles that were associated with active Mastodon accounts. We also investigated whether there exists any reciprocity in such Twitter-Mastodon profile associations; to this aim, we reverse-searched mentions of Twitter handles within the retrieved Mastodon user profiles, and found about 2.5K Mastodon profiles that declared a Twitter handle in their metadata, of which 98% correspond to a valid match between the mentioned handle and the original corresponding Twitter account (refer to Methods for the detailed procedure).

We analyzed the networks of social links between these 75K users on both Twitter and Mastodon. The Twitter follower network contains approximately 4M links, whereas the Mastodon network exhibits a relatively lower count of 2.5M links, representing a reduction of approximately 38%. The differences observed at the macroscopic structural level between the two networks are primarily due to their contrasting number of edges. In comparison to Twitter, the Mastodon network exhibits relatively lower density, lower average degree, lower transitivity, and an increased presence of small disconnected components. The fewer connections observed in the Mastodon network could potentially indicate that, at the time of data gathering, the process of link formation among newly-migrated users was still underway. The complementary cumulative distribution functions of the in-degrees in the two networks follow a similar trend that however diverges in a relatively large regime (approximately from 10 up to around $10^4$, as shown in Figure 1, left). Despite the variation in degree distribution and edge density, both networks exhibit a similarly high clustering coefficient, modularity, and percentage of reciprocal follower links. These shared characteristics suggest that both networks foster tightly-knit, reciprocal local neighborhoods that are arranged in well-separated communities.

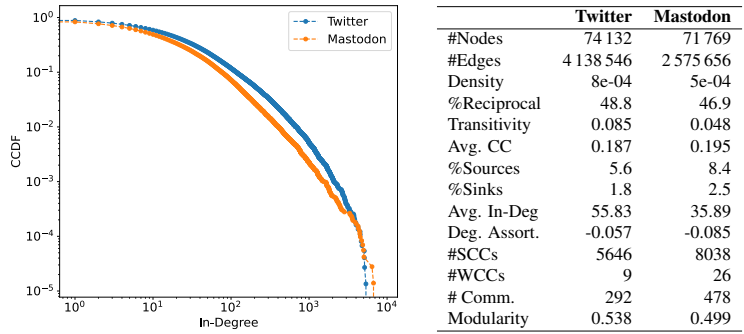

**Figure 1.** (Left) Comparison between the CCDF of the in-degree distributions of the Twitter and Mastodon networks. (Right) Main structural traits of the Twitter migration network and Mastodon network; the difference in the number of nodes is due to the presence of isolated nodes, i.e., migrated users having no ties with other migrated users in the Mastodon network.

The connections within the Mastodon network mirror in part the social relationships observed in the Twitter network. Approximately 30% of the links observed in Twitter can be found within Mastodon. When focusing on the *backbone* of the Twitter network, which is a pruned graph whereby spurious connections are filtered out (see Methods), the proportion of Twitter links replicated on Mastodon increases to 41%. This significant presence of shared edges, particularly in the backbone network, suggests that the migration was in part driven by the desire to "replicate" the existing Twitter social network on a new platform, rather than establishing an entirely new social context; indeed, this appears to moderately hold at node-neighborhood level, as hinted by a Spearman correlation of 0.6 between the rankings of local clustering coefficients of shared nodes in Twitter and Mastodon graphs. The structural metrics of both networks are summarized in Figure 1 (right).

**Social Influence of Migrants**

The widespread practice of using Twitter to announce one's decision to migrate to Mastodon raises the question of whether social influence played a role in Twitter users' migration choices. To investigate this, we resorted to *compartmental epidemic models* to characterize the "infectiousness" of migration decisions. Epidemic models have been extensively used to simulate information diffusion within social systems,[16,42] under the principle that the process of influence spreads through social connections, akin to the transmission of communicable diseases through social interactions, and that the population under study is partitioned into predefined compartments expressing epidemiological states. This allows us to focus on understanding global patterns, not on "who-infects-whom", thus in contrast to the approach underlying stochastic information diffusion and maximization models[43] which assume the availability of a network of connections among individuals, possibly with additional information about user-attributes.[44]

In this study, we employed the widely known SIRS model, where the entire population is initially susceptible ($S$) to a disease, subsequently certain individuals become infected ($I$) and can transmit the disease to others through their social connections, and over time, infected individuals recover ($R$) and eventually become susceptible to re-infection. We also experimented with the SIR model, a simpler version that does not consider re-infections, obtaining similar results (see Supplementary Information).

We simulated the diffusion process with a daily granularity. On any given day $t$, the set of susceptible individuals ($S$) contains Twitter users who have not yet created a Mastodon account by day $t$. The set of infected individuals ($I$) comprises Twitter users who registered on Mastodon up until day $t$. We hypothesize that the triggers of social influence were the public announcements of Mastodon handles on Twitter. Under this assumption, re-infections could correspond to users who announced multiple times their commitment to migrating. However, it was impossible to collect precise information regarding the timing of such announcements for all users. To approximate the day of the announcement, we used the day when their Mastodon account was created. This approximation generally proved to be accurate, considering the tendency of users in advertising their Mastodon profiles in a short time following their registration (0 days as mode, 1 day as median, 5 days as mean) we observed for the subset of users ($N = 41K$) with both timestamps available (see Supplementary Information). The set $I$ also includes 4K Twitter users who had registered on Mastodon during 2022, prior to our data collection period, who could then have helped to foster the #TwitterMigration movement. These can be regarded as "early-migrant" Mastodon users, with an average registration time of 168 days before October $26^{th}$ 2022, and median registration time of 183 days, corresponding to the date Musk struck the deal to acquire Twitter. The remaining set of recovered individuals ($R$) was determined by the model (see Methods). $R$ corresponds to the set of Twitter users who have migrated to Mastodon but whose influence effect has worn out (i.e., their intention to migrate has vanished from their followers' timelines).

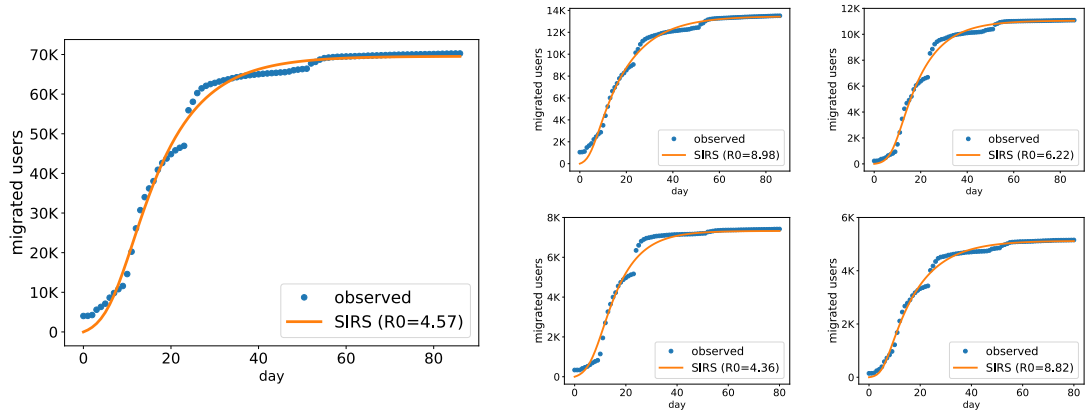

**Figure 2.** (Left) Cumulative number of Twitter users migrated to Mastodon over the course of 3 months since Elon Musk's acquisition of Twitter. Fit estimated with the SIRS model, along with its $R_0$. (Right) Fitting with the SIRS model for the top-4 largest communities by Louvain, reported in descending order row-wise.

To investigate whether the migration phenomenon was compatible with the characteristics of information diffusion, we tested whether a SIRS model could replicate the observed trend of the cumulative number of migrated users. To achieve this, we determined the parameters of the SIRS model that produced the best approximation of the empirical data (see Methods). The model takes the population size as input, and here we present results based on a population equivalent to the set of all migrated users at the end of our data collection period. In Supplementary Information, we provide additional results considering a larger population, equal to the size of users who have engaged in discussions involving Mastodon, including those who did not migrate in the timeframe of our study. The model with optimal parameters provided a close approximation of the empirical data, with a mean absolute percentage error (MAPE) of 0.077. From the parameters of the fitted model, we derived the *reproduction number* $R_0 = \beta/\gamma$, expressed as the ratio between the rate of infections $\beta$ and the rate of recovery $\gamma$. The $\beta$ parameter denotes the number of people with whom one infected individual interacts in a unit of time, whereas the parameter $\gamma$ models the number of people who recover in a unit of time. When $R_0 > 1$, the diffusion process grows as individuals become infected at a higher rate than they recover. Our models estimated an $R_0$ value of 4.57 (Figure 2, left), signifying a highly infectious process that backs our hypothesis of social influence in the migration process.

**Is Spreading Community-Driven?**

The Twitter follower network is highly modular (Figure 1, right), indicating that users participating in the #TwitterMigration discourse belong to distinct and loosely connected communities. This observation aligns with previous research highlighting the highly segregated nature of Twitter communities.[45] The presence of community structure plays a crucial role in information diffusion dynamics. Within communities characterized by dense social ties, information can rapidly spread among members, whereas community boundaries restrict the propagation of information to the rest of the social network.[46,47] We studied how the social influence process differed across communities.

To this aim, we first used the Louvain method to partition the Twitter follower network connecting migrated users into communities (see Methods), and obtained 292 well-separated communities (modularity of 0.538). The resulting communities are heterogeneous in size, with a long tail of very small isolated groups (see Figure S3 in Supplementary Information). We focused our analysis on the 44 communities containing a sufficiently high number of members (size $\geq 50$), that jointly account for 98% of nodes in the migrated network.

We fit the SIRS model on each of these communities, treating them as isolated systems. The estimated reproduction number across communities exhibits a range of values from $R_0 = 1$ to $R_0 = 11.82$, indicating that the process of social influence unfolded at varying rates depending on the social context. Figure 2 displays the fits and corresponding $R_0$ values for the four largest communities (refer to the Supplementary Information for a more exhaustive account on communities).

Notably, despite these fittings involve a different number of users and various rates of social influence, they appear to exhibit a similar trend. This common trait can be attributed to exogenous factors that have contributed to shape the course of migration. A key case in point is the restart of the diffusion process during the third week, prompted by Elon Musk's request to his employees to sign a pledge to work harder at the development of "Twitter 2.0" or leave with three months of severance pay[48].

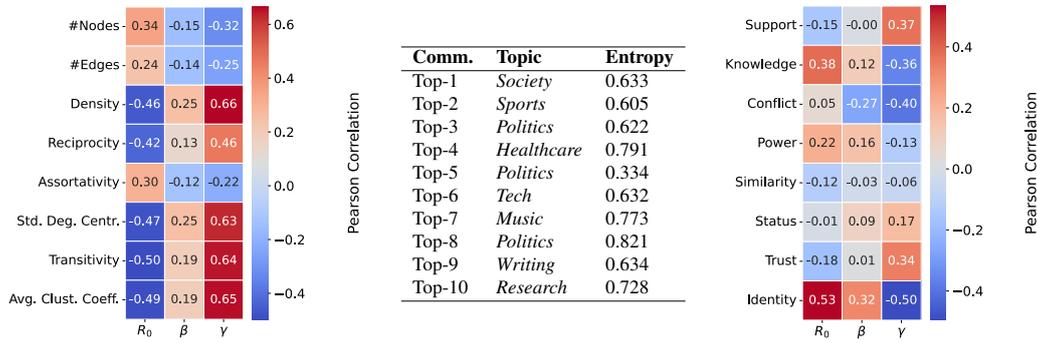

**Figure 3.** (Left/Right) Pearson correlation heatmap between topological and social aspects characterizing communities and main parameters of the SIRS model. (Center) Main topic of each of the top-10 largest communities with corresponding community-level topical entropy value. Each topic was manually labeled after excluding the inherent Twitter and Mastodon discourse, which would have hidden the real topics.

We also noticed that in most communities, the SIRS model slightly improves on the SIR model in terms of fit with the empirical data, with an average MAPE of 0.124 compared to 0.125 by the SIR; moreover, higher reproduction numbers occur in SIRS, with an average $R_0$ of 5.08 against 3.92 in SIR (see Supplementary Information), which highlights the role that repeated signals of commitment, which are inherently captured by the SIRS model, may have played in the social influence process.

To gain insights into the factors that contribute to the acceleration of the influence process, we explored the correlations between the parameters of the fitted SIRS models in different communities and three categories of factors that have previously been associated with the adoption of new opinions and behaviors in social groups,[19] namely *network topology*, *reiterated commitment*, and *language use*. We elaborate on each of these aspects next.

**Network features.** In terms of network topology, we investigated various community-specific structural aspects, including network size, density, reciprocity, distribution of prestige, and different connectivity metrics (Figure 3, left). Notably, we found only a weak correlation between community size and the basic reproduction number (Pearson correlation $r = 0.343$). Communities with higher values of $R_0$ tend to exhibit sparser connectivity ($r = -0.459$ w.r.t. density), as well as lower levels of reciprocity ($r = -0.425$). Furthermore, communities that experienced faster diffusion were characterized by social ties linking users with similar levels of prestige ($r = 0.298$ w.r.t. assortativity), by a less pronounced hierarchical structure, where the prestige of individuals (measured by their number of followers) was more evenly distributed ($r = -0.474$ w.r.t. standard deviation of in-degree centrality), and by lower levels of clustering ($r = -0.497$ w.r.t. transitivity, and $r = -0.488$ w.r.t. average clustering coefficient).

**Reiterated commitment.** We examined the influence of iterated commitment on the migration process in terms of the tweets concerning the #TwitterMigration within each of the identified communities. To this aim, we collected a dataset of $\sim 8.3M$ tweets posted by migrated users during the initial month of the #TwitterMigration. We intentionally excluded tweets posted on December 2022, due to their prevalent shifted focus on commentaries related to the 2022 FIFA World Cup.

First, to provide an indicator of the extent to which Mastodon discussions are prevalent within each community, we calculated the *commitment* as the ratio $n_M/n_{tot}$ between the number of tweets about Mastodon ($n_M$) and the total volume of tweets ($n_{tot}$) posted by community members over our examination period. We found a positive correlation ($r = 0.54$) between the commitment and the value of $R_0$ measured at community level, suggesting that a greater and reiterated focus on Mastodon in community discussions might have aided the migration process.

Second, we carried out topic modeling of our tweet corpus, by means of BERTopic, a topic modeling method that has been shown to be more effective in extracting topics from short texts like tweets compared to other traditional topic modeling techniques like Latent Dirichlet Allocation[49]. Based on the resulting users' topic distributions, for each community we selected the distributions of its members, calculated the entropy of such topic distributions, and correlated it with the corresponding $R_0$ of that community. We noticed a significant variety of entropy values over all communities, which also moderately correlates with $R_0$ ($r = 0.20$), suggesting that the discussion of a broadest range of topics might lead to higher infectiousness. In addition, we spotted that most communities are characterized by mid-high entropy values, which is surprisingly negatively correlated with the community sizes ($r = -0.23$). An overview of the most prominent topics in the ten largest communities, along with their corresponding entropy scores, is shown in Figure 3 (center). The above findings hint that *(i)* the migratory process involves users discussing multiple topics, and *(ii)* users migrating might be further persuaded when they perceive broad topical variety.

|                         | **Predicting $R_0$ from:** |                         |
|-------------------------|----------------------------|-------------------------|
| **Topological features** | **Activity features**    | **Social Features**    |

| Feature | β | SE | p | Feature | β | SE | p | Feature | β | SE | p |
|---|---|---|---|---|---|---|---|---|---|---|---|
| Density | -0.459 | 0.137 | 0.002 | Commitment | 0.542 | 0.130 | 0.000 | Knowledge | 0.394 | 0.117 | 0.002 |
| | | | | | | | | Identity | 0.543 | 0.117 | 0.000 |
| Durbin-Watson stat. = 2.639 | $R^2_{adj}$ = **0.192** | | | Durbin-Watson stat. = 1.990 | $R^2_{adj}$ = **0.277** | | | Durbin-Watson stat. = 1.837 | $R^2_{adj}$ = **0.412** | | |

| **Topological & Activity Features** | | | | **Social & Activity Features** | | | | **Topological & Social & Activity Features** | | | |
|---|---|---|---|---|---|---|---|---|---|---|---|
| Feature | β | SE | p | Feature | β | SE | p | Feature | β | SE | p |
| Density | -0.406 | 0.116 | 0.001 | Knowledge | 0.333 | 0.117 | 0.007 | Density | -0.356 | 0.104 | 0.002 |
| Commitment | 0.499 | 0.116 | 0.000 | Identity | 0.408 | 0.131 | 0.003 | Knowledge | 0.340 | 0.104 | 0.002 |
| | | | | Commitment | 0.271 | 0.134 | 0.050 | Identity | 0.305 | 0.120 | 0.015 |
| | | | | | | | | Commitment | 0.282 | 0.119 | 0.023 |
| Durbin-Watson stat. = 2.473 | $R^2_{adj}$ = **0.430** | | | Durbin-Watson stat. = 1.878 | $R^2_{adj}$ = **0.453** | | | Durbin-Watson stat. = 2.302 | $R^2_{adj}$ = **0.568** | | |

**Table 1.** Ordinary Least Squares regression model fittings for the prediction of $R_0$ from topological, activity, and social features (top row), and their combinations (bottom row). β coefficients describe the contribution of each feature to the outcome, along with the standard errors (SE) and statistical significance (p-values). Auto-correlation is evaluated via the Durbin-Watson statistic (values closest to 2 indicate no auto-correlation). Regression results are reported via adjusted $R^2$.

**Language use.** Finally, we delved into the pragmatics of language use in relation to the migration process. Specifically, we focused on the *social pragmatics* of language, namely the intended social function of an utterance. Prior research in the social sciences has extensively explored the associations between various forms of social intent and the process of opinion formation. For instance, conveying *trust* has been identified as a crucial factor in aligning divergent points of view.[50] Recent studies have surveyed a range of *dimensions* of social pragmatics commonly observed in everyday language,[51] and have developed language models capable of classifying conversational text based on these dimensions.[26] In the context of Reddit, such a model has been employed to examine the relationship between the use of specific dimensions in public argumentation and an increased likelihood of opinion change.[52] We built a *social dimensions classifier* (see Methods) to our dataset of tweets to identify those that, with high probability, conveyed *Support*, *Knowledge*, *Conflict*, *Power*, *Similarity*, *Status*, *Trust*, and *Identity*. We then computed the ratio $C_{i,d} = N_{i,d}/\mathcal{N}_{i,d}$, which represents the number of tweets marked with dimension $d$ in community $i$ ($N_{i,d}$) divided by the expected number of tweets conveying dimension $d$ in the same community if tweets with dimension $d$ were uniformly distributed at random throughout the entire network ($\mathcal{N}_{i,d}$). Community-level $R_0$ exhibited the strongest correlation with the ratios of *Knowledge* ($r = 0.38$) and *Identity* ($r = 0.53$) (Figure 3, right). These correlations suggest that information exchange and, particularly, the sense of belonging to a shared group or community might have played key roles in the migration phenomenon.

**Drivers of Migration**

We observed that factors related to network *topology*, individual *commitment*, and *language use* correlate individually with the speed of the social influence across Twitter communities. To explore the interplay among these three aspects, we conducted experiments using various regression models to predict community-specific $R_0$ based on combinations of features that displayed the strongest correlations (see Table 1). All variables were standardized before using them in the regressions. A more exhaustive set of regressions is presented in Supplementary Information. A model that solely incorporates network *density* yielded the poorest fit ($R^2_{adj} = 0.192$, β = -0.459). When considering commitment alone, a slightly higher correlation was observed ($R^2_{adj} = 0.277$). When combining density and commitment, the goodness of fit approximately doubled ($R^2_{adj} = 0.430$), indicating that these two factors jointly account for over 40% of the variability in $R_0$ across communities. Interestingly, a comparable level of fit was achieved by solely considering the prevalence of knowledge and identity messages ($R^2_{adj} = 0.412$). Ultimately, the best fit was obtained by combining all variables ($R^2_{adj} = 0.568$). In this last model, all predictors maintain statistical significance ($p < 0.05$), and the magnitude of their coefficients is comparable, meaning that each variable contributes non-negligible signals in modeling the susceptibility of communities to behavioral changes. Lower density of social links and abundant exchange of factual knowledge exhibit slightly stronger associations with $R_0$ in this multivariate model, followed by expressions of identity and, last, by iterated expressions of commitment.

## Discussion

In pursuit of experimental evidence of factors underlying behavioral change in social communities, we have studied the #TwitterMigration phenomenon: a rapid and extensive migration of Twitter users to the decentralized social platform Mastodon. Few other studies have touched upon this event, and have done so from angles that are different from our own. They either focused on the decentralization properties of the Mastodon ecosystem from the perspective of the migrated users,[14] or explored the characteristics of Mastodon communities that are associated with higher rates of user retention.[15] Those studies were conducted on data that was either smaller or with fewer dimensions (e.g., no analysis of text) than what we consider in this work.

Our study makes a first attempt at describing the dynamics of this migration from an information diffusion perspective, finding that a simple epidemic model of information spreading closely replicates the temporal trace of migration. Crucially, we observed that the effectiveness of the social influence to migrate (i.e., the value of $R_0$ estimated by the epidemic model) was community-dependent, and found patterns that help explain why some communities were more successful in migrating more rapidly. Drawing inspiration from prior research on behavioral change, we tested three hypotheses pertaining to the *structure* of the social network, the *commitment* of community members, and their *language use* as potential explanations for these observed differences. By testing each of these hypotheses, we gained interesting insights.

The first finding of our study is that the only structural factor that was consistently associated with an increased $R_0$ is the sparsity of social connections. This is somewhat unexpected, because it contradicts the conventional understanding that close social proximity, characterized by high clustering and dense social connections, leads to faster and broader adoption of new behaviors.[53] We propose two non-exclusive explanations for this counter-intuitive trend. Firstly, the discussion surrounding the #TwitterMigration phenomenon competed with numerous other topics vying for the limited attention of Twitter users. The prominence of #TwitterMigration in the overall discourse might have been more diluted in denser follower networks.[54] Secondly, the incentive to migrate may simply be proportional to the fraction of a user's friends who have already migrated, and such fraction increases more rapidly in networks with fewer social connections.

Our second finding is that communities engaging in more frequent discussions on the #TwitterMigration topic exhibited higher $R_0$. We interpret this result in the light of theories linking the rapid emergence of consensus to the influence exerted by committed individuals, even when they constitute a small minority.[19] While this intuition is grounded in strong theoretical foundations,[20,55] empirical support has been limited.[56] Our operationalization of commitment, as a simple and straightforward measure to quantify the level of engagement in discussing a specific topic, contributes to validating the underlying theoretical framework.

The third and final finding of our study is that communities in which the influence process unfolded more rapidly are those whose discourse frequently emphasizes a shared *identity* and engages in substantial exchanges of factual *knowledge*. Extensive research in social psychology has established a connection between psycho-linguistic aspects of social interaction and successful, spontaneous coordination.[22–25] In particular, the Identity Theory posits that cooperation can be facilitated through cognitive mechanisms that foster a sense of belonging to the same social group,[57] suggesting that identity may be pivotal in overcoming social dilemmas involving coordinated behavior that entail inherent risks or a non-zero cost of action.[58] Moreover, the exchange of truthful, factual information has been identified as a prerequisite for constructive debates and, ultimately, persuasion.[59] The combined influence of identity and knowledge explains more variance in our data than the combined influence of density and commitment, highlighting the significant role of psycho-linguistic aspects as key drivers of behavioral change.

Our work comes with limitations that future work can address.

First, our perspective of the social system in which the migration occurred is limited in several ways. Specifically: *(i)* it is plausible that the migration process continued beyond the temporal boundaries captured by our dataset; *(ii)* We focused solely on the network of users who explicitly disclosed their migration, not accounting for users who might have migrated silently and not considering any potential influences from other regions of the Twitter network or exogenous events such as news items ; and *(iii)* we treated the follower network as a comprehensive proxy for all the information channels available to our users on Twitter, which is not true in general. It is hard to gain complete knowledge on the set of Twitter users who migrated to Mastodon without employing more costly information gathering techniques such as extensive surveys. To provide evidence on the robustness of our results to the choice of the set of migrated users, we modeled network influence in two extreme scenarios: one in which we consider only the social graph that connects the 75K users who declared their migration, and one in which the social graph includes all users who were involved in the Mastodon discussion (described in Supplementary Material). Crucially, we observed that the temporal patterns of migration in both scenarios are well described by a process of information diffusion with relatively high contagiousness, represented by $R_0$ values exceeding one. We argue that a scenario in which one could obtain the full list of migrated users would lie in between these two extremes, and thus still exhibit relatively high values of $R_0$.

We also conjecture that "silent" users who migrated without explicitly signaling their migration might have not considerably contributed to spreading social influence, effectively making them akin to non-migrated users from the perspective of the information diffusion phenomenon.

Second, our measurements are inherently limited in terms of precision, scope, and validity: *(i)* we adopted only a macroscopic perspective on the migration phenomenon, relying on measurements derived from population-level or community-level aggregates; *(ii)* the network communities identified through unsupervised techniques may not accurately reflect the communities as subjectively perceived by Twitter users; and *(iii)* both supervised (social dimensions) and unsupervised (topic modeling) methods employed in the analysis of natural language are prone to errors — while they provide valuable insights when interpreted in aggregate, they may fail to accurately categorize specific instances of text.

Last, the factors we explored in relation to the migration rate are not exhaustive, and collectively account for only slightly more than half of the observed variance. The vast literature on behavior change, social influence, and collective action encompasses a broader array of factors than those considered in our study. We encourage future research to investigate additional elements beyond the scope of our work, in order to gain a more comprehensive understanding of the complex dynamics at play.

## Methods

### Data Collection

**Migration-related Tweets.** No public datasets about the #TwitterMigration movement was available at the time of writing. To fill this gap, we carried out an extensive data crawl on both Twitter and Mastodon. We used the full-archive search functionality of Twitter's Academic API v2 to download tweets relevant to the migration that were posted from October $26^{th}$ 2022 (when Twitter acquisition by Musk was finalized), to January $19^{th}$ 2023. First, to identify hashtags that were frequently mentioned in the discourse about the migration, we started from the list of hashtags that were featured as *trending* in the days following the public announcement of the acquisition. We then expanded those hashtags by snowball sampling: we collected all the tweets containing those trending hashtags published between October $26^{th}$ 2022 and November $26^{th}$ 2022, and counted the frequency of all hashtags mentioned in those tweets. We then manually parsed the top hashtags in the frequency distribution and identified a set of 13 hashtags that unambiguously referred to the migration. To ensure high recall to our data, our dataset is comprised of tweets that *(i)* contain one of the 13 hashtags we found by snowball expansions, or *(ii)* mention Mastodon's official Twitter account (i.e., @joinmastodon), or *(iii)* contain the keyword "mastodon". We excluded retweets. The final query we submitted to the Twitter API was: " (@joinmastodon OR mastodon OR #TwitterMigration OR #RIPTwitter OR #TwitterTakeover OR #TwitterShutdown OR #TwitterIsDead OR #MastodonSocial OR #LeavingTwitter OR #ElonIsDestroyingTwitter OR #MastodonMigration OR #Fediverse OR #Mastodon OR #TwitterAlternative OR #DecentralizedSocialMedia) -is:retweet ". Together with the tweets, we saved the full profile description of the user who posted the tweet, as returned by the APIs.

**Matching Twitter Handles with Mastodon Handles.** In an effort to recreate their Twitter social network on Mastodon, some users promoted their Mastodon *handles* on Twitter. We used this information to link Twitter users with their corresponding Mastodon profiles. To achieve this, we first devised different regular expressions to identify potential Mastodon-like handles occurring either in the *username*, *description*, or *tweets* of each user. We found $108k$ handles that were compatible with Mastodon profiles. Further refinement of the matching strings was necessary due to the prevalent format of Mastodon handles, which aligns with that of email addresses (i.e., username@domain.tld). To distinguish strings specifically referring to Mastodon accounts, we cross-referenced them with a compiled list of known Mastodon instances that we obtained from the `instances.social` APIs, the most widely-recognized and comprehensive tracker of Mastodon instances. This step filtered the set of handles down to ∼75K. Last, for each of the Mastodon handles we found, we queried the official Mastodon API (`https://docs.joinmastodon.org/api/`) to collect *(i)* the list of all their followers and followees, and *(ii)* their profile metadata, including the `created_at` field, which records the precise timestamp of account creation. To comply with general privacy-preserving policies that might be set forth by Mastodon instances, we did not acquire any textual or multimedia content.

### Network Modeling

**Twitter and Mastodon Social Graphs.** We built the social contexts of each migrant user on the two platforms as follows. We first collected the followees of all ∼ 75K migrated Twitter users, obtaining ∼ 111M raw links between ∼ 16.6M users, and analogously we collected ∼ 20M raw links involving the set of followees of migrated users on Mastodon.

We then modeled two distinct social graphs from the Twitter and Mastodon data we collected. The first graph, denoted as $\mathscr{G} = \langle V, E, t \rangle$, is a directed and node-labeled graph, which we also refer to as the *migration network*. In fact, its vertex set $V$ represents the ∼ 75K Twitter users who have a corresponding Mastodon account in our dataset, and the set of edges $E$ models

∼ 4M social ties through the follower relationship, with $(i, j) \in E$ indicating that user $i \in V$ follows user $j \in V$. Each user is also labeled with its migration date through the function $t : V \mapsto T$, which assign a node in $V$ with a timestamp in $T$ denoting the creation date of the corresponding user's Mastodon account. The second graph, denoted as $\mathscr{G}_M = \langle V_M, E_M \rangle$, represents the directed and unweighted Mastodon follower network. It comprises a set $V_M$ of Mastodon users that we collected, and a set $E_M$ of 2.5M edges that represent the follower relationships among these Mastodon users, where $(i, j) \in E_M$ denotes that Mastodon user $i \in V_M$ follows Mastodon user $j \in V_M$.

**Graph Backboning.** Given the noisiness of online social ties, it is appropriate to perform a network simplification or backboning task aimed at detecting and pruning noisy edges or ties formed due to random chance, thus bringing out the latent structure of a network. In this regard, albeit the simplest solution might be exploiting edge weights to filter out all edges having a weight below a fixed and pre-determined global threshold, the risk of removing social ties locally relevant yet weak at the network level is evident. To this aim, we resorted to the *Disparity Filter* (*DF*)[60] method, which exploits *generative null model* based on node distribution properties to prune networks from statistically irrelevant edges, i.e., via p-value computation w.r.t. specific significance levels. Specifically, the *DF* leverages the null hypothesis that the strength of a node is redistributed uniformly at random over the node's incident edges, thus evaluating the strength and degree of each node locally. By using its publicly available implementation (https://github.com/malcolmvr/backbone_network), we applied *DF* to our migration network equipped with an edge weighing function defined as follows. Each edge in $\mathscr{G}$ is assigned a weight through a function $w : E \mapsto \mathbb{R}$, which measures for each $(i, j) \in E$ the similarity between $i$ and $j$ as the Jaccard coefficient over the out-neighbors of $i$ and the in-neighbors of $j$, i.e., $i$ following $j$ is regarded as similar to $j$ proportionally to as much users followed by $i$ tend to follow $j$, thus reflecting a simple notion of prestige conferred by $i$ to $j$.

**Community Detection.** To unveil the underlying community structure within our *migration network*, we employed the widely-adopted Louvain algorithm for community detection.[61] We used its original (undirected) implementation due to its exceptional scalability that enables accurate community detection in large-scale networks. The Louvain algorithm adopts a hierarchical greedy approach to maximize the modularity $Q$, defined as:

$$Q = \frac{1}{2m} \sum_{i,j} \left[ A_{ij} - \frac{k_i k_j}{2m} \right] \delta(c_i, c_j), \tag{1}$$

where $A_{ij}$ represents an entry in the binary adjacency matrix between nodes, $k_i$ and $k_j$ denote the degrees of nodes $i$ and $j$, respectively, and $m$ corresponds to the total number of edges. The Kronecker delta function $\delta(c_i, c_j)$ takes the value 1 when nodes $i$ and $j$ belong to the same community ($c_i = c_j$), and 0 otherwise. Overall, the modularity $Q$ denotes the quality of a network partitions into communities by weighting the density of intra-community links against inter-community links; it ranges between -0.5 and +1, with higher values indicating better partitioning.

The Louvain algorithm starts by assigning each node to a singleton community and then proceeds by merging communities aiming at maximizing the modularity gain $\Delta Q$, that for a node $i$ is defined as:

$$\Delta Q = \left[ \frac{\sum_{in} + 2k_{i,in}}{2m} - \left( \frac{\sum_{tot} + k_i}{2m} \right)^2 \right] - \left[ \frac{\sum_{in}}{2m} - \left( \frac{\sum_{tot}}{2m} \right)^2 - \left( \frac{k_i}{2m} \right)^2 \right] \tag{2}$$

In the equation, $\sum_{in}$ is the sum of links within the target community, $\sum_{tot}$ is the sum of links incident to nodes in the target community, $k_i$ is the sum of links incident to node $i$, $k_{i,in}$ is the sum of links from node $i$ to nodes in the target community, and $m$ is the total number of edges in the network. The algorithm stops when no more improvements in modularity can be achieved.

**Epidemiological Models.** To verify whether the temporal pattern of migration is compatible with an information diffusion phenomenon, we adopted the hypothesis of *infodemic spreading*[16] and fit the SIR and SIRS *compartmental epidemiological models*[62] to our data. The SIR model divides the population into three compartments: *susceptible* (S), *infectious* (I), and *recovered* (R) individuals. The model operates under three assumptions: *(i)* a closed world assumption where individuals cannot enter or leave the population during the spreading phenomenon, *(ii)* equal susceptibility of all individuals to the information, and *(iii)* once recovered, individuals cannot be reinfected. The SIR model is described by the following set of differential equations:

$$\begin{aligned} dS/dt &= -\beta S \cdot I/N \\ dI/dt &= \beta S \cdot I/N - \gamma I \\ dR/dt &= \gamma I \end{aligned} \tag{3}$$

In the formula, $S$, $I$, and $R$ represent the number of susceptible, infectious, and recovered individuals, respectively. $\beta$ denotes the infection rate, $\gamma$ represents the recovery rate, and $N$ indicates the total population size. The basic reproduction number, $R_0 = \beta/\gamma$, serves as a measure of the average number of infections generated by an infected individual, indicating the speed of contagion. In the context of our study, an $R_0 > 1$ corresponds to a scenario of a growing infodemic in which, over time, more and more users decide to migrate.

The SIRS model extends SIR by allowing previously recovered individuals to return to a susceptible state. This model introduces the loss-of-immunity rate parameter, $\xi$, indicating the rate at which individuals lose their immunity and become susceptible again. The updated set of differential equations is as follows:

$$\begin{aligned} dS/dt &= -\beta S \cdot I/N + \xi R \\ dI/dt &= \beta S \cdot I/N - \gamma I \\ dR/dt &= \gamma I - \xi R \end{aligned} \qquad (4)$$

In our study, a reinfection event may signify users reaffirming their commitment to migrating. We estimated the parameters of both models using the least square estimation method,[63] which involves minimizing the sum of the squares of the residuals, computed as the difference between the observed data depicting the spreading of the infection and the predictions by the corresponding model. We generated these predictions by solving the systems of Ordinary Differential Equations (ODEs) underlying our compartmental models as an initial value problem, fed with information involving the onset of the outbreak such as the overall population and the initially infected individuals, and leaving the model finding the set of remaining parameters that would minimize the residuals, thus providing the best fit for the observations.

**Network Analysis Measures.** To characterize the main topological traits underlying the networks in our study, we resorted to classic structural measures, which are summarized as follows. *Transitivity* and *Average Local Clustering Coefficient* provide a global and local perspective of triadic closure, respectively, where the former refers to the probability that two incident edges are completed by a third one to form a triangle, and the latter expresses how much connected are the neighbors of a node, averaged over all nodes. *Source* and *Sink* nodes denote nodes with no incoming, resp. no outgoing, links and provide insights about the origin and termination of information flows within a network. The *Degree Assortativity* measures the tendency of nodes to link with peers having similar degrees, and in this work corresponds to the Pearson correlation between the out-degree of source nodes and the in-degree of target ones. *Strongly* and *Weakly Connected Components* inform about the connectivity of a network in terms of isolated subgraphs; the weakly case refers to subnetworks in which every node can be reached from every other node by ignoring edge directionality, whereas the strongly case requires taking into account the latter. The *Standard Deviation* of the *Degree Centrality* provides insights into the heterogeneity of importance (based on the number of connections nodes have) within a network, where higher values suggest a greater disparity, with some nodes being more relevant than others.

## Language Modeling

**Topic Modeling.** To extract the latent topics that characterize the tweets of a user, we employed BERTopic,[64] a recently-developed unsupervised topic model that leverages Transformer-based pre-trained language models to extract coherent and descriptive topics from unstructured text. The BERTopic pipeline consists of three stages. First, it transforms documents into a high-dimensional embedding space that captures the semantics of the input text. Second, it reduces the embedding to a lower-dimension space that is more suitable for clustering. Lastly, it applies clustering algorithms to identify groups of documents corresponding to distinct topics.

Given a time window of interest, we shape the social discourse of each migrated user by creating a document corresponding to the concatenation of all tweets posted during such a period. We then fed such documents to BERTopic to infer a representative topic for each user based on their tweets. Based on such information, we also computed the topical entropy for each community in our Twitter graph, to quantify the diversity of the collection of topics discussed by the users of that community.

We tried different settings of the main hyperparameters for the various stages of BERTopic. Here we report the best performing configuration. For generating embeddings, we used the default Sentence-BERT,[65] specifically its pre-trained *all-mpnet-base-v2* model. To reduce the dimensionality of the embeddings, we employed Principal Component Analysis (PCA) to project them into a 5-dimensional space. Subsequently, we carried out the K-Means algorithm with $n_{clusters} = 120$ to cluster the reduced embeddings. We used a *CountVectorizer* and a *c-TF-IDF* modules to extract accurate topic representations. Specifically, to improve topic quality, we set $stop\_words = english$ and $ngram\_range = (1,3)$ for the former and $reduce\_frequents = True$ for the latter, respectively. Finally, we fine-tuned topics by using the *MaximalMarginalRelevance* module with $diversity = 0.5$ to avoid similar keywords in our topics and the *KeyBERTInspired* module to improve the semantic relationship between keywords and documents in each topic. Other optimizations involve the specification of the minimum size of a topic should have ($min\_topic\_size = 175$) and the automatic topic reduction after training the model ($nr\_topics = auto$).

It should be noted that a systematic exploration of BERTopic hyperparameters goes beyond the objectives of this work; rather, we chose to keep all the components in the BERTopic pipeline, and experimented with a few variations of the pipeline according to the recommended best-practices in using BERTopic, primarily focusing on finding an appropriate setting to achieve a reasonable trade-off between meaningfulness of the induced topics and training/inference time speedup.

**Social Dimensions.** To infer the social intent that Twitter messages convey, we resort to a theoretical model of ten social dimensions that reflect fundamental social aspects of social interactions. These dimensions have been identified through surveys and an extensive literature review, and they are frequently expressed in social conversations, for example to give emotional support or to convey trust.[51] Previous work developed a model[26] that can accurately detect the presence of these dimensions from conversational text, and that has been successfully employed in multiple studies of online conversations.[52,66–68] We used the publicly available LSTM[69] implementation of the model (http://www.github.com/lajello/tendimensions) that has been trained on around 9k manually labeled sentences, and demonstrated good performance, achieving an average Area Under the Curve (AUC) of 0.84 across the various dimensions.[26]

Building upon the idea that textual components might be associated with a combination of the ten dimensions[51], instead of being a multi-class classifier, the underlying model comprises a set of independently-trained classifiers, one for each dimension, functioning as a multi-label classifier. Given a message $m$, each classifier estimates a score $s_d(t) \in [0,1]$ for each sentence $t$ in $m$, that encodes the likelihood that the sentence conveys the target dimension; then it returns the maximum score across all sentences $s_d(m) = max_{t \in m} s_d(t)$, thus allowing individual sentences to convey a given dimension and avoiding affecting results by averaging low-score sentences. This intuition matches with theoretical interpretations[51] according to which also brief expressions can unveil dimensions.

To ease the interpretation of the results, we binarize the scores $s_d(m)$ to split messages between those that carry dimension $d$ with high probability and those that do not. We do so through an indicator function that assigns a specific dimension $d$ to a given message $m$ if its corresponding score is above a pre-defined threshold $\theta_d$.

To account for the dimension-dependent empirical distribution of the classifier scores, we avoided relying on a fixed common threshold, and we rather defined dimension-specific thresholds. We adopted a conservative, high-precision approach, and set $\theta_d$ to the $90^{th}$ percentile w.r.t. the empirical distribution of the scores $s_d$, thus effectively narrowing the set of messages marked with each dimension down to the 10% of the total messages.

## Acknowledgements

LMA acknowledges the support from the Carlsberg Foundation through the COCOONS project (CF21-0432). AT is partly funded by the PNRR FAIR project (H23C22000860006, M4C21.3 spoke 9). The funders had no role in study design, data collection and analysis, decision to publish, or preparation of the manuscript.

## Author contributions

All authors conceived the experiments and wrote the manuscript. LLC collected the data and analysed the results.

## Competing interests

The authors declare no competing interests.

# Supplementary Information for "Drivers of social influence in the Twitter migration to Mastodon"

**Lucio La Cava**[1,*], **Luca Maria Aiello**[2,3], **and Andrea Tagarelli**[1]

[1]University of Calabria, Rende, Italy
[2]IT University of Copenhagen, Denmark
[3]Pioneer Centre for AI, Denmark
[*]lucio.lacava@dimes.unical.it

## S1.1 Development and signaling of the migration process

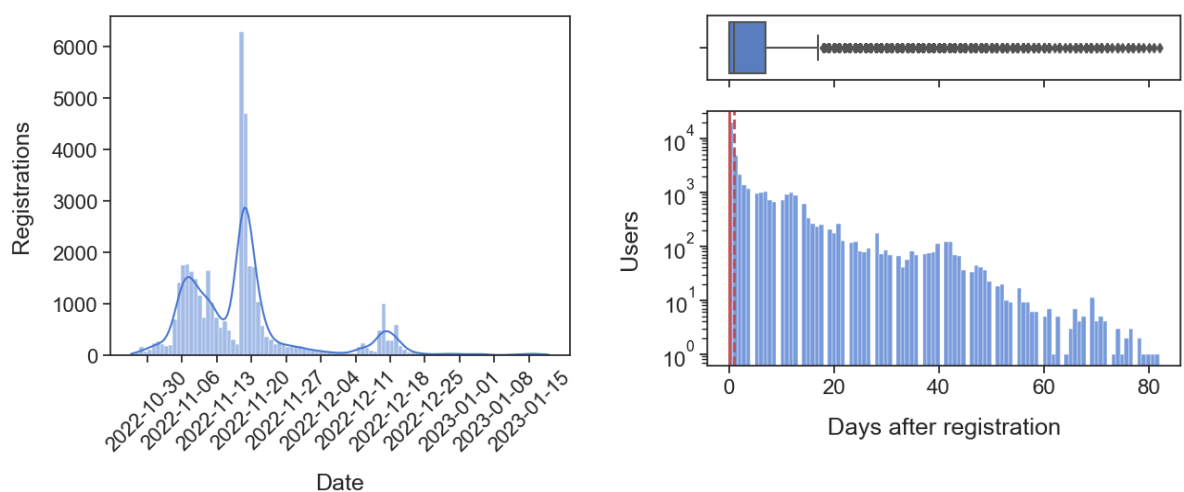

**Figure S1.** (Left) Registration dates of users migrated to Mastodon between October 26$^{th}$ 2022 and January 19$^{th}$ 2023. (Right) Time delta between migration to Mastodon and signaling of the new Mastodon handles on Twitter; the solid and dashed red lines represent the mode and median of the distribution, respectively.

We analyzed how the #TwitterMigration movement developed as a result of users discussing on Twitter their intention to switch to Mastodon. We hypothesizes that users announcing their Mastodon profiles on Twitter acted as a social trigger that might have persuaded other users to migrate. Given that the date of such announcement is not available for all users, we used the timestamp of Mastodon account creation as a proxy. Figure S1 (left) reports the distribution of registration times after the purchase of Twitter by Elon Musk. Interestingly, some spikes in registrations emerge in response to controversial choices by the new management (e.g., mass layoffs and policy changes). Nonetheless, for our proxy hypothesis to be valid, it is important that the registration time is close to the time in which the Mastodon account was announced on Twitter. We could measure the temporal gap between these two events for a subset of about 41$K$ users for whom we have both the registration time on the new platform and the signaling time of the corresponding profile on Twitter. As shown in Figure S1 (right), the time gap is rather short for most users, with mode of 0 days (i.e., less than 24 hours), a median of 1 day, and an average of 5 days.

## S1.2 Fitting with the SIR model and with an extended set of users related to the #TwitterMigration

**Fitting the SIR model.** In addition to the SIRS model, we experimented with the simpler SIR model. We observed that the SIR model could replicate the observed trend of the cumulative number of migrated users equally (same $R_0$ and MAPE values) with the SIRS model, as reported in Figure S2 (left). However, in this regard we point out that although this may seemingly suggest that from a macroscopic point of view both models perform identically, this turns out to be a borderline case, as in general the SIRS model has been shown to be superior in terms of goodness of fit (cf. Results) due to its ability to model reiterated commitment in our infectiousness scenario.

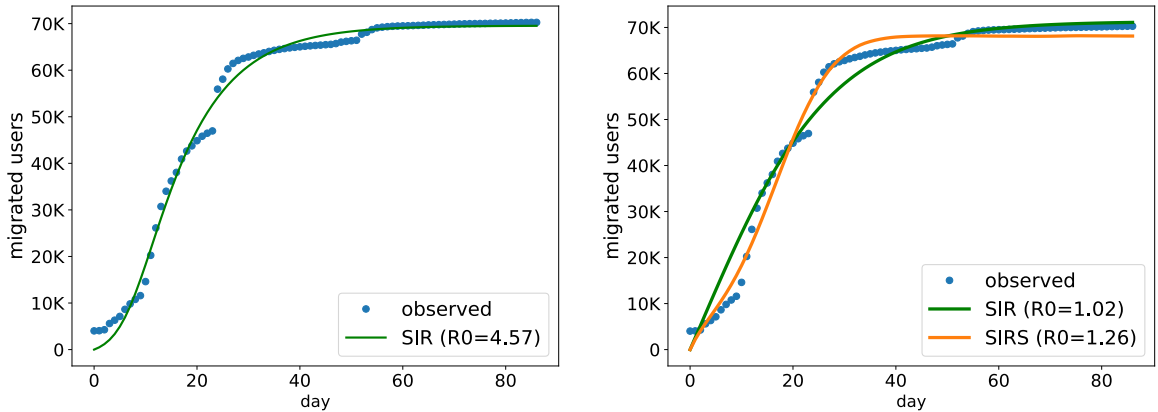

**Figure S2.** Cumulative number of Twitter users migrated to Mastodon over the course of 3 months since Elon Musk's acquisition of Twitter. (Left) Fit estimated with the SIR compartmental epidemiological model. (Right) Fit estimated with the SIR and SIRS compartmental model with the set of users corresponding to the population including all users who have discussed Mastodon on Twitter.

**Expanding the population.** Our primary focus in this work is on the analysis of the main drivers of social influence behind the #TwitterMigration movement, and accordingly we considered as the population underlying our epidemic models the set of users eventually migrated to Mastodon. Nonetheless, to conduct a comprehensive study, we extended our evaluation by broadening the set of users representing our population. Specifically, this expanded set includes $\sim$ 540K users who have engaged in discussions involving Mastodon between October $26^{th}$ 2022 and January $19^{th}$ 2023, regardless of their final decision to migrate or not. Although this choice provides an underestimation of the actual process due to the inclusion of potential noise in the population, our SIR and SIRS models estimated an $R_0$ value of 1.02 and 1.26 (see Figure S2, right), respectively. Since $R_0$ is greater than 1, this confirms growth of the infection process underlying social influence.

## S1.3 Further details on communities

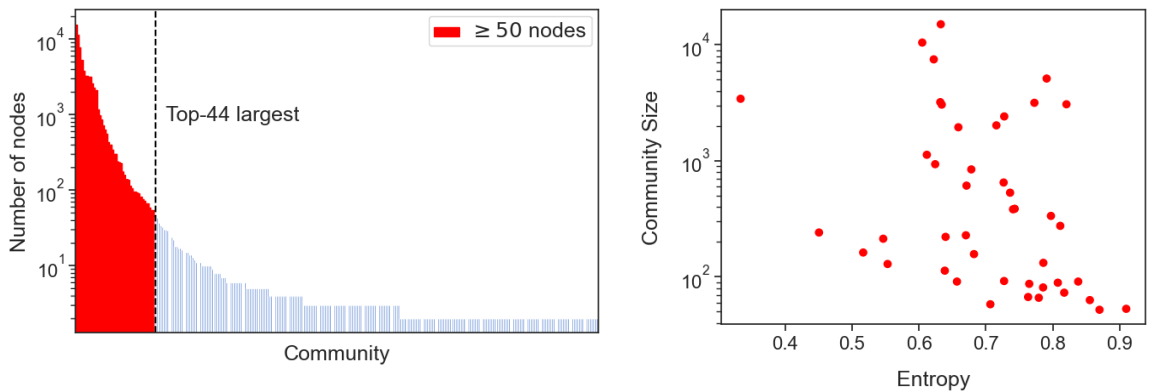

**Figure S3.** (Left) Distribution of users across the communities found via the Louvain method; red-colored communities contain at least 50 users. (Right) Scatterplot showing the relationship between community topical entropy and size for the top-44 largest communities.

To investigate the spreading of the #TwitterMigration's infectiousness across communities, we focused our analysis on a specific group of nodes containing a sufficiently representative number of users. Specifically, by referring to the distribution reported in Figure S3 (left), it is possible to observe that most users turn out to be clustered in a few communities, with a long tail of communities containing a few members, or even singleton nodes. Consequently, for our analysis, we selected the set of communities having at least 50 users, which gives us coverage of 98% of the set of migrated users, corresponding to the top-44 largest communities. In this regard, we also complement our findings on the relationship between community size and topical entropy by resorting to Figure S3 (right), which shows the (moderate) negative correlation we spotted between these two quantities across the top-44 largest communities.

## S1.4 Fitting compartmental models with the largest communities

Figure S4 shows how the SIR and SIRS model fit the migration data for the top-15 largest communities by number of users. In line with our findings (cf. Results), both the SIRS and SIR models are able to accurately shape the migration process, albeit with different $R_0$ (with mean values of 5.08 for SIRS and 3.92 for SIR for the top-44 largest communities).

## S1.5 Experimenting with regression models to predict community-specific $R_0$

Table S1 reports the regression model fittings for predicting community-specific $R_0$ based on the full set of topological and social features we leveraged in our analyses. Table S2 reports the fitting of a Least Absolute Shrinkage and Selection Operator (LASSO) regression model on the combination of the best-performing features for predicting $R_0$ (cf. Results).

| **Predicting $R_0$ from:** | | | | **Predicting $R_0$ from:** | | | |
|---|---|---|---|---|---|---|---|
| **Topological Features** | | | | **Social Features** | | | |
| **Feature** | **β** | **SE** | **p** | **Feature** | **β** | **SE** | **p** |
| Density | -0.055 | 0.967 | 0.955 | Support | 0.040 | 0.188 | 0.834 |
| Reciprocity | -0.123 | 0.235 | 0.604 | Knowledge | 0.507 | 0.161 | 0.003 |
| Assortativity | -0.015 | 0.270 | 0.955 | Conflict | -0.043 | 0.171 | 0.802 |
| Std. Deg. Centr. | 0.230 | 1.345 | 0.865 | Power | 0.248 | 0.157 | 0.124 |
| Transitivity | -0.426 | 0.604 | 0.485 | Similarity | -0.103 | 0.220 | 0.641 |
| Avg. Clust. Coeff. | -0.175 | 0.500 | 0.729 | Status | -0.126 | 0.194 | 0.519 |
| | | | | Trust | 0.098 | 0.153 | 0.526 |
| | | | | Identity | 0.511 | 0.141 | 0.001 |
| Durbin-Watson stat. = 2.661 | | $R^2_{adj}$ = **0.140** | | Durbin-Watson stat. = 2.038 | | $R^2_{adj}$ = **0.395** | |

**Table S1.** Ordinary Least Squares regression model fittings for the prediction of $R_0$ from the full set of topological (left) and social (right) features. β coefficients describe the contribution of each feature to the outcome, along with the standard errors (SE) and statistical significance (p-values). Auto-correlation is evaluated via the Durbin-Watson statistic (values closest to 2 indicate no auto-correlation). Regression results are reported via adjusted $R^2$.

| **Predicting $R_0$ from:** | |
|---|---|
| **Topological & Social & Activity Features** | |
| **Feature** | **β** |
| Density | -0.274 |
| Knowledge | 0.244 |
| Identity | 0.246 |
| Commitment | 0.240 |
| $R^2_{adj}$ = **0.525** | |

**Table S2.** Least Absolute Shrinkage and Selection Operator regression model fitting for the prediction of $R_0$ from a combination of topological, activity, and social features. β coefficients describe the contribution of each feature to the outcome. Penalty term is set to 0.1. Regression results are reported via adjusted $R^2$.

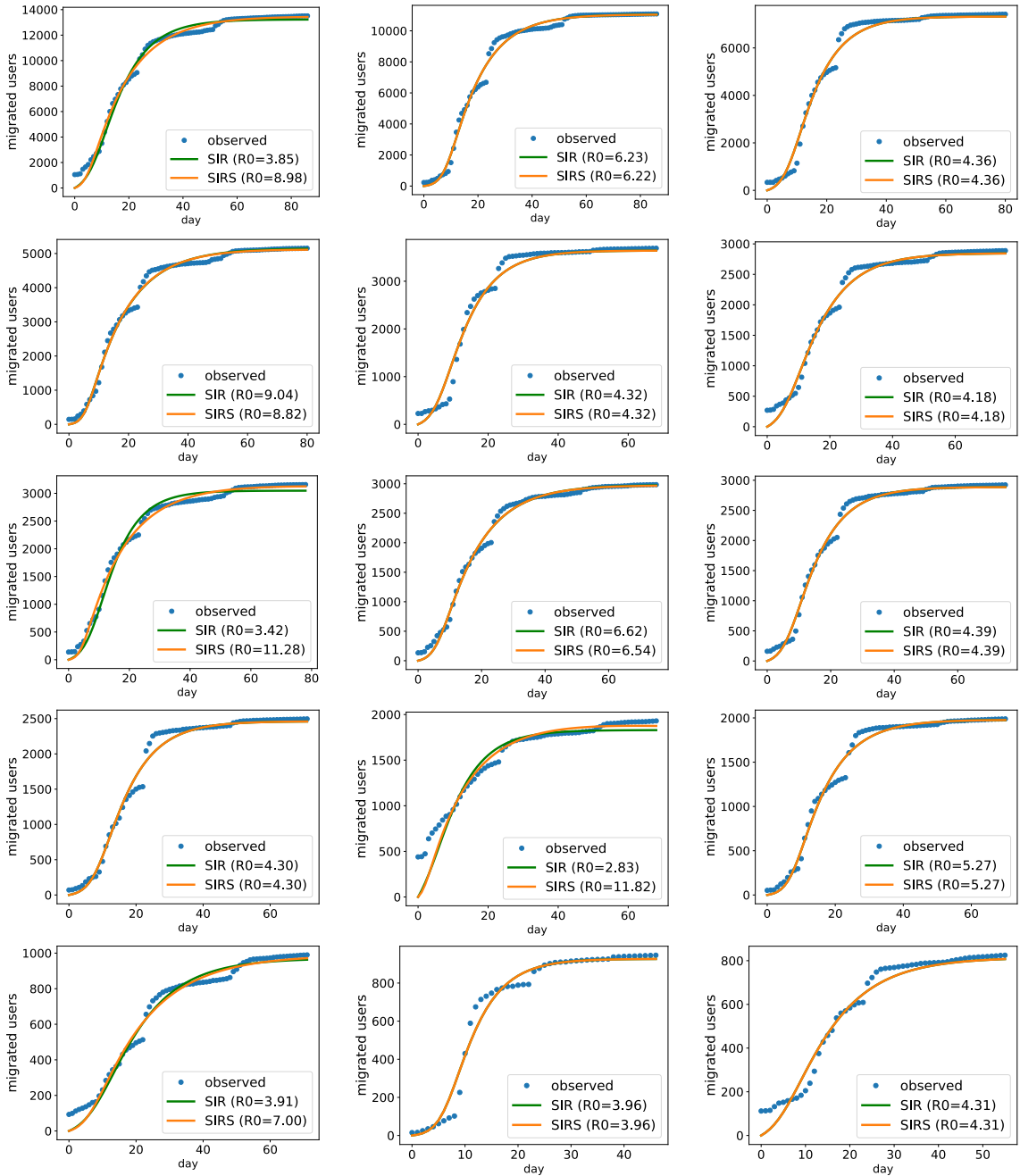

**Figure S4.** Fitting with the SIR and SIRS compartmental epidemiological models and associated $R_0$ values for the top-15 largest communities by Louvain, reported in descending order row-wise, of the cumulative number of Twitter users migrated to Mastodon over the course of 3 months since Elon Musk's acquisition of Twitter.

## S1.6 User activity in the months following the migration

To assess the effectiveness of the migration from Twitter to Mastodon almost a year after the Musk's buyout, we examined the activity levels of migrated users. To this aim, we utilized Mastodon's official APIs to retrieve the date of the most recent post made by migrated users. This check was performed in mid-October 2023. The presence of a valid payload not only confirmed the user's ongoing activity (i.e., post creation) on Mastodon, yet also allowed us to determine its extent. Figure S5 (left) summarizes the key findings from our analysis. Remarkably, more than 95% of the migrated users maintained their account on Mastodon. Furthermore, while we observed a decline in activity due to the initial surge following the collective migration, we found that almost half of the migrated users remained active six months after the migration, suggesting interest in well-rooted settlement from a considerable fraction of migrated users. We took a closer look at users who made particular efforts to remain active even after a substantial period following the surge in migrations. To this aim, we calculated the fraction of users who continued to contribute by creating content in the last few months. Notably, as reported in Figure S5 (left), more than a quarter of migrated users were observed to be actively posting during the month leading up to our mid-October 2023 check. These users constitute a resilient segment of the migrated population that was able to endure the initial noisy growth due to the curiosity to try out a new platform, thus effectively managing to cave out a new social space and maintaining their engagement and participation in the community. We further delved into such an investigation by computing, for each migrated user $i$, the corresponding *persistency* and *freshness* levels. The former is defined as $p_l(i) = t_p(i)/t_m(i)$ and indicates the duration of which the user $i$ has been active (considering the creation of posts) on the platform $t_p(i)$, compared to the time has passed since migrating $t_m(i)$. The latter has been defined in previous work[1] as $f_l(i) = 1/\log_2(2 + t_\Delta)$, where $t_\Delta$ indicates the number of days elapsed between the date of checking the user's activity (i.e., mid-October 2023) and the date of the last post created by the user. These two complementary scores were normalized in $[0,1]$ and serve as proxies for the migrated users' degree of persistency and propensity to contribute, respectively. As reported in Figure S5 (right), migrated users almost split in a bipartite fashion. Indeed, while we observed a moderate fraction of migrated users not actively engaging in the new platform, most of them were found to be clustered at high persistency levels, with observable peaks around the maximum value. Similarly, despite observing that a large fraction of migrated users are not particularly keen to contribute, we found two noticeable groups exhibiting mid and high values of freshness, respectively, thus actively contributing with fresh content on the new platform. We further evaluated the extent of this propensity to contribute, unveiling very intriguing traits. Indeed, as reported in Figure S5 (left), these "fresher" users were responsible for generating nearly the entire volume (90.13%) of posts by migrated users on the new platform. This is particularly evident among those in the last quartile of freshness value (73.56%), which hence act as "super-users". These intriguing findings underscore the pivotal role played by these super-users in sustaining the post-migration ecosystem and pave the way for further investigations. Finally, we explored the potential connection between the level of persistency/freshness and the underlying social influence in the migration process. Remarkably, we observed a non-negligible correlation between the $R_0$ value of the SIRS model and the corresponding persistency ($p = 0.491$) and freshness ($p = 0.394$) levels of the top-44 communities on which we narrowed our focus within our study. This finding, coupled with the replication of more than 40% of the social ties after the migration (cf. Results), poses a stepping point for the proper understanding of the growing migratory phenomena between social platforms, thereby warranting further investigations.

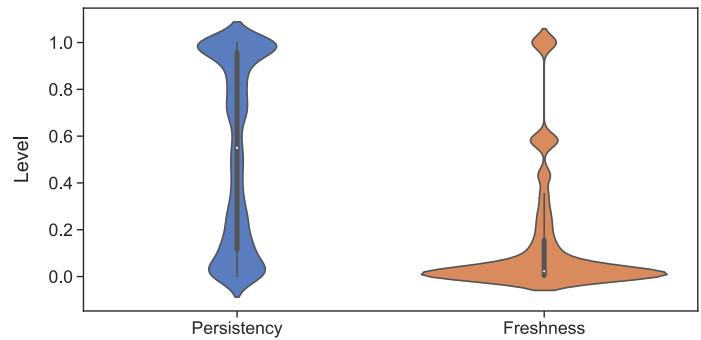

| Check | Perc. |
|---|---|
| Account existence | 95.13% |
| Active after 1 month | 69.26% |
| Active after 3 months | 54.60% |
| Active after 6 months | 44.73% |
| Active in the last 3 months | 37.61% |
| Active in the last 2 months | 32.78% |
| Active in the last month | 27.50% |
| Volume $\geq$ Q3 freshness | 73.56% |
| Volume $\geq$ Q2 freshness | 90.13 % |

**Figure S5.** Percentage of accounts still existing and active w.r.t. different temporal checkpoints. (Right) Violin plots for the persistency and freshness levels of migrated users.